\documentclass[twocolumn,superscriptaddress,showkeys]{revtex4-1}

\usepackage{epsfig}
\usepackage{amsmath}
\usepackage{amssymb}
\usepackage{amsfonts}
\usepackage{mathptmx}
\usepackage{dcolumn}
\usepackage{eucal}
\usepackage{bm}
\usepackage{color}
\usepackage[colorlinks,linkcolor=blue,citecolor=blue]{hyperref}
\usepackage{epstopdf}

\begin{document}

\title{Electrostatic control of phase slips in Ti Josephson nanotransistors}

\author{C. Puglia}
\affiliation{Dipartimento di Fisica, Universit\`a di Pisa, Largo Bruno Pontecorvo 3, I-56127 Pisa, Italy}
\affiliation{NEST, Istituto Nanoscienze-CNR and Scuola Normale Superiore, I-56127 Pisa, Italy}
\author{G. De Simoni}
\email{giorgio.desimoni@sns.it}
\affiliation{NEST, Istituto Nanoscienze-CNR and Scuola Normale Superiore, I-56127 Pisa, Italy}
\author{F. Giazotto}
\email{francesco.giazotto@sns.it}
\affiliation{NEST, Istituto Nanoscienze-CNR and Scuola Normale Superiore, I-56127 Pisa, Italy}

\begin{abstract}
The investigation of the switching current probability distribution of a Josephson junction is a conventional tool to gain information on the phase slips dynamics as a function of the temperature. Here we adopt this well-established technique to probe the impact of an external static electric field on the occurrence of phase slips in gated all-metallic titanium (Ti) Josephson weak links. We show, in a temperature range between 20 mK and 420 mK, that the evolution of the phase slips dynamics as a function of the electrostatic field starkly differs from that observed as a function of the temperature. 
This fact demonstrates, on the one hand, that the electric field suppression of the critical current is not simply related to a conventional thermal-like quasiparticle overheating in the weak-link region. 
On the other hand, our results may open the way to operate an electrostatic-driven manipulation of phase slips in metallic Josephson nanojunctions, which can be pivotal for the control of decoherence in superconducting nanostructures.
\end{abstract}

\maketitle


\section{Introduction}
Although a static electric field is almost ineffective on the conduction properties of metals, recent experiments demonstrated the possibility to suppress via conventional gating the critical current $(I_C)$ of metallic Bardeen-Cooper-Schrieffer superconducting wires \cite{desimoni2018}, Dayem bridges \cite{likharev1979,paolucci2018,paolucci2019,Paolucci2019b}, and of proximity superconductor-normal metal-superconductor (SNS) Josephson junctions (JJ) \cite{desimoni2019}. Yet, by means of a superconducting quantum interference device (SQUID) consisting of two gated Dayem bridge constrictions, it has been possible to directly measure the impact of a static electric field on the  quantum phase difference $(\phi)$ across a Josephson weak-link \cite{Paolucci2019a}. 
The electric field was found to influence the SQUID current-phase relation via direct suppression of the critical current of a gated weak-link. In addition, unexpectedly, phase shifts in the SQUID current vs flux relation were measured for gate voltage values low enough to have no influence on $I_C$. Phase fluctuations present inside the superconductor \cite{Paolucci2019a} were shown to be a plausible cause for such an effect to occur, suggesting that the influence of the electric field on the phase may also lead to the occurrence of phase slips, i.e., local random 2$\pi$ jumps of $\phi$ \cite{langer1967}, which are responsible for the superconducting-to-normal state transition \cite{little1967}. 
Phase slip events, indeed, reflect into the value of the switching current ($I_S$), that is, when the bias current is swept from zero to above the critical current, the bias value at which the superconductor switches to the normal state. 
Due to the stochastic nature of phase slip events, $I_S$ statistically spreads around $I_C$, and its distribution [also known as the switching current probability distribution (SCPD)] naturally provides information on the phase slips dynamics of mesoscopic superconducting devices \cite{sahu2009,bezryadin2000,zaikin1997,kim2017,kim2017a,pekker2009,coskun2012,bae2012,murphy2015,baumans2017,murphy2013,foltyn2015,Zgirski2018,kivioja2005,foltyn2015,ejrnaes2019,shah2008,aref2012} under the influence of external parameters such as, for instance, the temperature or an externally applied electric field. 
The latter was recently investigated in hybrid graphene-based Josephson junctions \cite{lee2011,choi2013}, but no relationship between the electric field and the SCPDs has been observed  so far in genuine all-metallic superconducting systems. 
The relevance of such topic lies in its natural link with the study of decoherence mechanisms in JJ devices, a matter of strong interest mainly in view of the realization of advanced superconducting quantum information architectures. On this regard, we want to explicitly mention two superconducting qubit implementations that would benefit from a deep knowledge of the effect of electric fields on the dynamics of phase slippages in JJs: all-metallic transmons \cite{larsen2015} and phase-slip qubits \cite{mooij2005}. In the first case, the study of relation between electric field and SCPD is of great relevance because it allows to identify and control one of the sources of phase decoherence related to the manipulation of the qubit state via field-effect. In the phase slip qubit case, the controlled generation of phase slips in a JJ is at the basis of the manipulation mechanism of the qubits. Therefore, the possibility to regulate the phase slip rate, via a control electrode, would provide a convenient knob for operating on a phase-slip qubit. We wish also to highlight that, the control
of the phase slips in a JJ is a tool to reduce, in binary-logic superconducting devices (see, e. g. reference \cite{paolucci2019}), the number of the errors due to the
random transitions from the superconducting to the normal state.
\begin{figure*}[ht!]
\centering
\includegraphics[width=\textwidth]{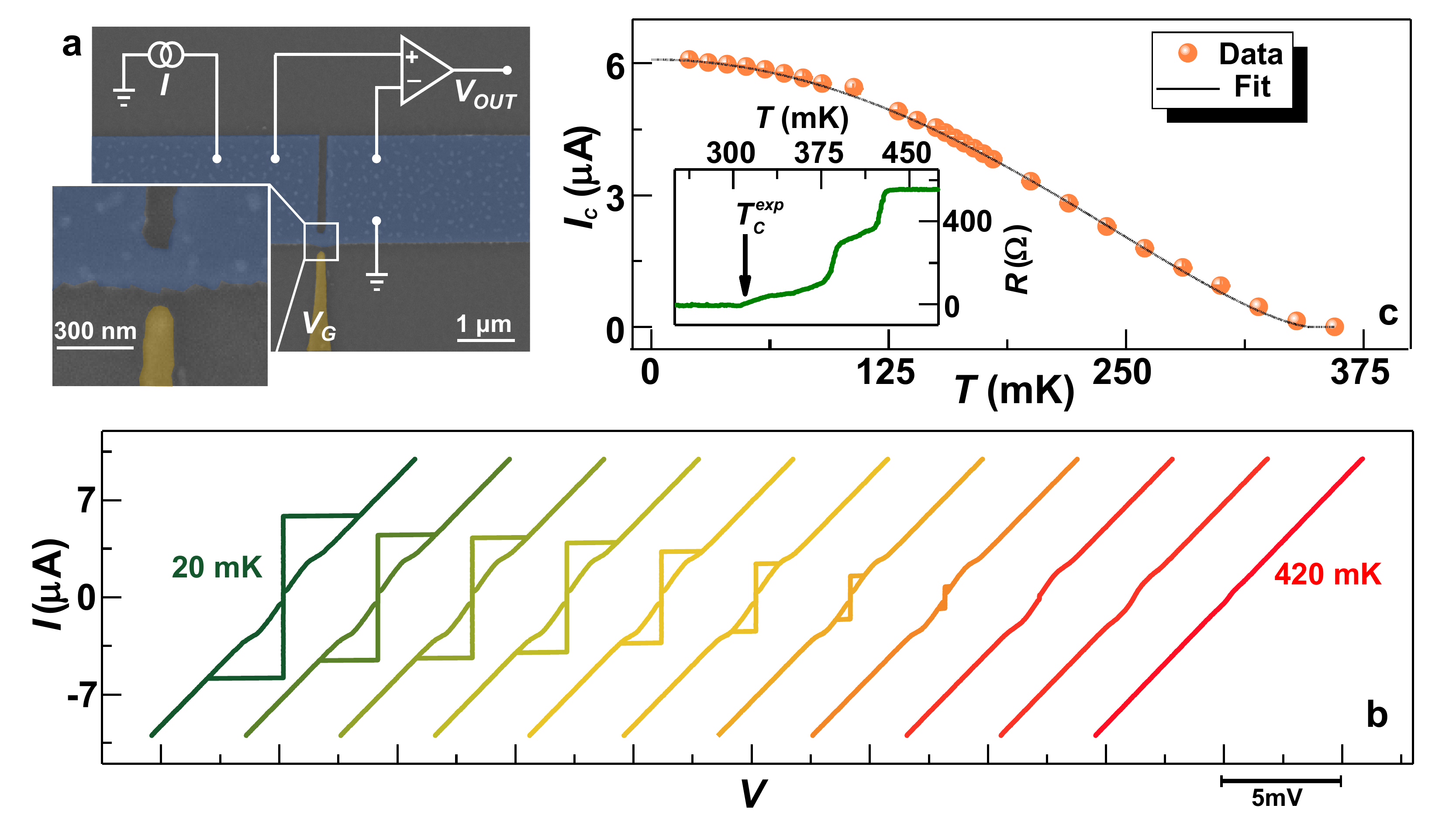}
\caption{(a) False-color electron micrograph of a typical Ti Dayem bridge Josephson transistor. The Josephson weak link (inset) is current biased, and the voltage drop is measured with a room-temperature voltage pre-amplifier, while the gate voltage ($V_G$) is applied to a side gate electrode (yellow). 
(b) Back and forth current $I$ vs voltage $V$ characteristics of a representative device measured at different bath temperatures from 20 mK to 420 mK in steps of 40 mK. The curves are horizontally offset for clarity.
(c) Evolution of the critical current $I_C$ as a function of the temperature (dots). The dashed line represents the evolution of $I_C$ according to Bardeen's theory \citep{bardeen1962}. 
The error bars on the measurement of $I_C$, calculated as the 
 standard deviation $\sigma$ of $I_S$ over $10^4$ samplings, is smaller than the dots size in this scale. 
The inset shows the weak-link resistance ($R$) as a function of the bath temperature $T$. The estimated critical temperature ($T_C^{exp}\sim 310\,$mK) is indicated by an arrow.}
\label{fig:fig1}
\end{figure*}

Here we tackle the point of understanding the link between the application of a gate voltage and the occurrence of phase slippages in JJ, and report the investigation of SCPDs in electrostatically-controlled titanium Dayem bridge Josephson weak-links in a temperature range from 20 mK to 420 mK, a regime explored so far only for Josephson tunnel junctions \cite{blackburn2014}.  
Our analysis of SCPDs of gated Dayem bridges JJs demonstrates the dramatic action of the electrostatic field on the phase slips dynamics in metallic superconductors, and opens the way to operate an electric field-driven control of phase slips and, thereby, of decoherence in Josephson weak-links. Moreover, we will show that the evolution of SCPDs as a function of the electrostatic field starkly differs from that measured as a function of temperature. 
This fact indicates that the electric field-driven critical current suppression is not related to a mere thermal-like quasiparticle overheating occurring in the junction region \cite{Timofeev2009}.

\begin{figure*}[ht!]
\centering
\includegraphics[width=\textwidth]{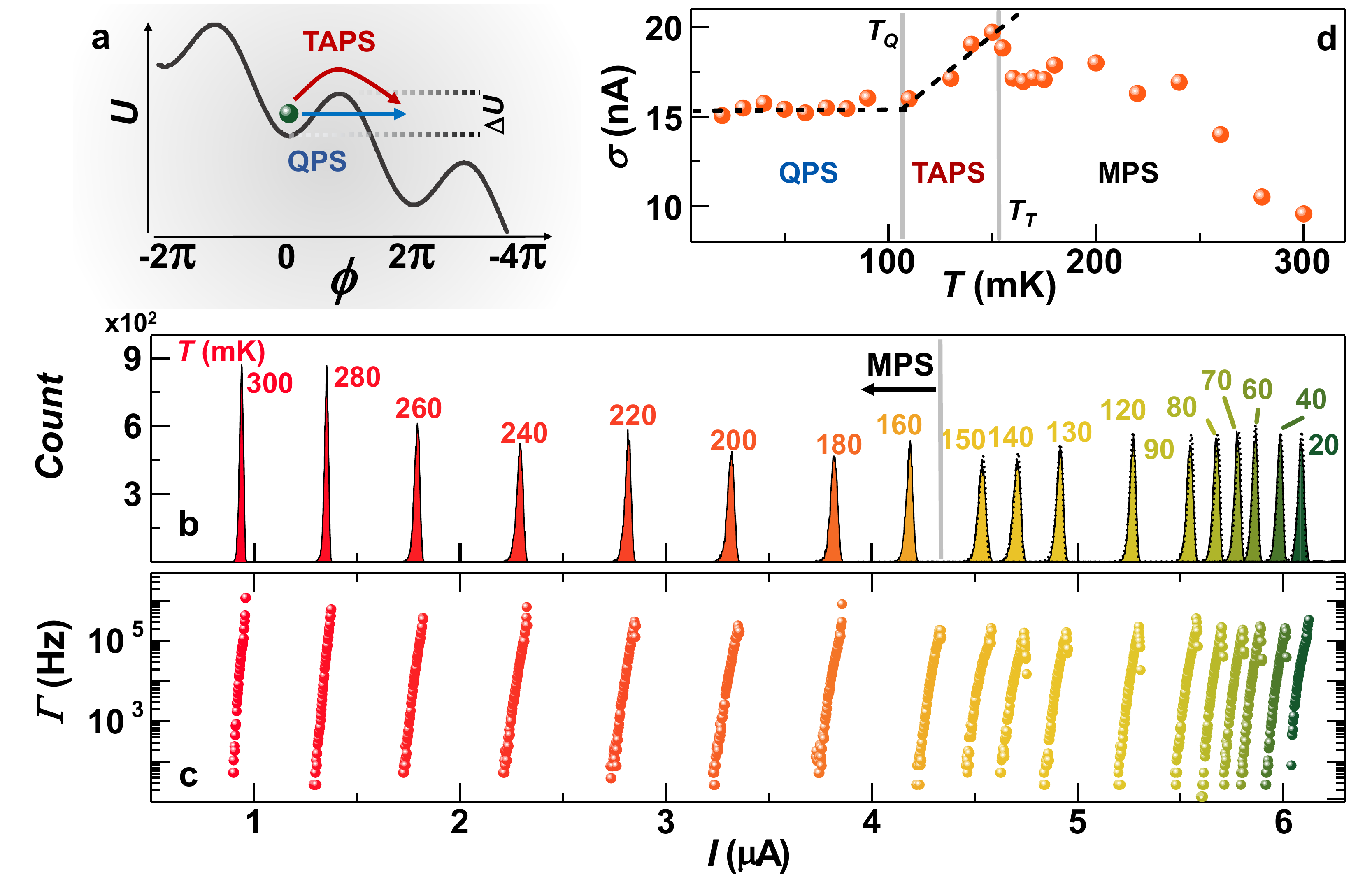}
\caption{(a) Schematic diagram of the tilted washboard potential showing the quantum phase slip (QPS, blue arrow) and thermally-activated phase slip (TAPS, red arrow) processes. $\Delta U$ is the height of the barrier defined as the energy difference between a minimum and the following maximum of the washboard potential (black line). 
(b) Switching current probability distributions (SCPDs) vs current $I$ obtained at different temperatures from 20 mK to 300 mK.  Dotted lines represent the best-fit curves obtained with KFD model. The gray vertical line indicates the crossover temperature from TAPS to MPS regime. 
(c) Experimental rate $\Gamma$ vs $I$ obtained with the direct KFD transform for the same temperature values as in panel (b). 
(d) Standard deviation $\sigma$ of the SCPDs vs bath temperature $T$. The crossover temperatures, $T_Q\simeq 110$ mK and $T_T\simeq 160$ mK, separate QPS/TAPS and TAPS/MPS regimes, respectively.
For each SCPD the total sampling number of $I_S$ is $10^4$.}
\label{fig:fig2}
\end{figure*}
 
\section{Experimental}
\subsection{Device nano-fabrication}
Our Ti-based Dayem bridges weak-links consist of 30-nm-tick, 150-nm-long, 120-nm-wide planar gated junctions fabricated by a single-step electron beam lithography of a poly methyl methacrylate (PMMA) resist mask deposited onto a sapphire $($Al$_2$O$_3)$ single-crystal wafer with a nominal resistivity larger than $10^{10}$ $\Omega\cdot $cm. Titanium was evaporated at a rate of 1.2 nm/s in an ultra-high vacuum electron-beam evaporator with a base pressure of about $\sim 10^{-11}$ Torr. The 140-nm-wide gate electrode was separated by a distance of about 80 nm from the Dayem bridge constriction. 
Figure \ref{fig:fig1}a shows the false color scanning electron micrograph of a representative Josephson device.

\subsection{Low-temperature preliminar electric characterization}
The low-temperature electric characterization of the devices was obtained by standard dc four-wire current versus voltage $(I$ vs $V)$ technique in a filtered cryogen-free $^3$He$-^4$He dilution refrigerator, carried out with a low-noise current generator and room-temperature differential voltage pre-amplifier. 
Figure \ref{fig:fig1}b shows the back and forth $I$ vs $V$ characteristics of a typical JJ device registered at several bath temperatures. The curves exhibit the conventional hysteretical behaviour, which stems from heating induced in the weak-link when switching from the dissipative  to the dissipationless regime \cite{courtois2008} (the device normal-state resistance is $R_N\simeq 550\,\Omega$). $I_C$ decreases with temperature according to the behaviour expected from the Bardeen's formula \cite{bardeen1962} $I_C (T)=I_C^0\left[1-\left(\frac{T}{T_C}\right)^2\right]^\frac{3}{2}$, where $I_C^0$ is the zero-temperature critical current, and $T_C$ is the critical temperature of the superconducting weak-link. The fit of the $I_C$ vs $T$ characteristic with Bardeen's equation (shown as the black line in Fig. \ref{fig:fig1}c) yields $I_C^{0\ (fit)}\simeq 6.02 \mu$A and $T_C^{(fit)}\simeq348$ mK. The latter value is in reasonable agreement with the critical temperature $T_C^{(exp)}\simeq310$ mK extracted from the low-frequency lock-in resistance $(R)$ versus $T$ measurement (see the inset of Fig. \ref{fig:fig1}c).  
\begin{figure*}[t!]
\centering
\includegraphics[width=\textwidth]{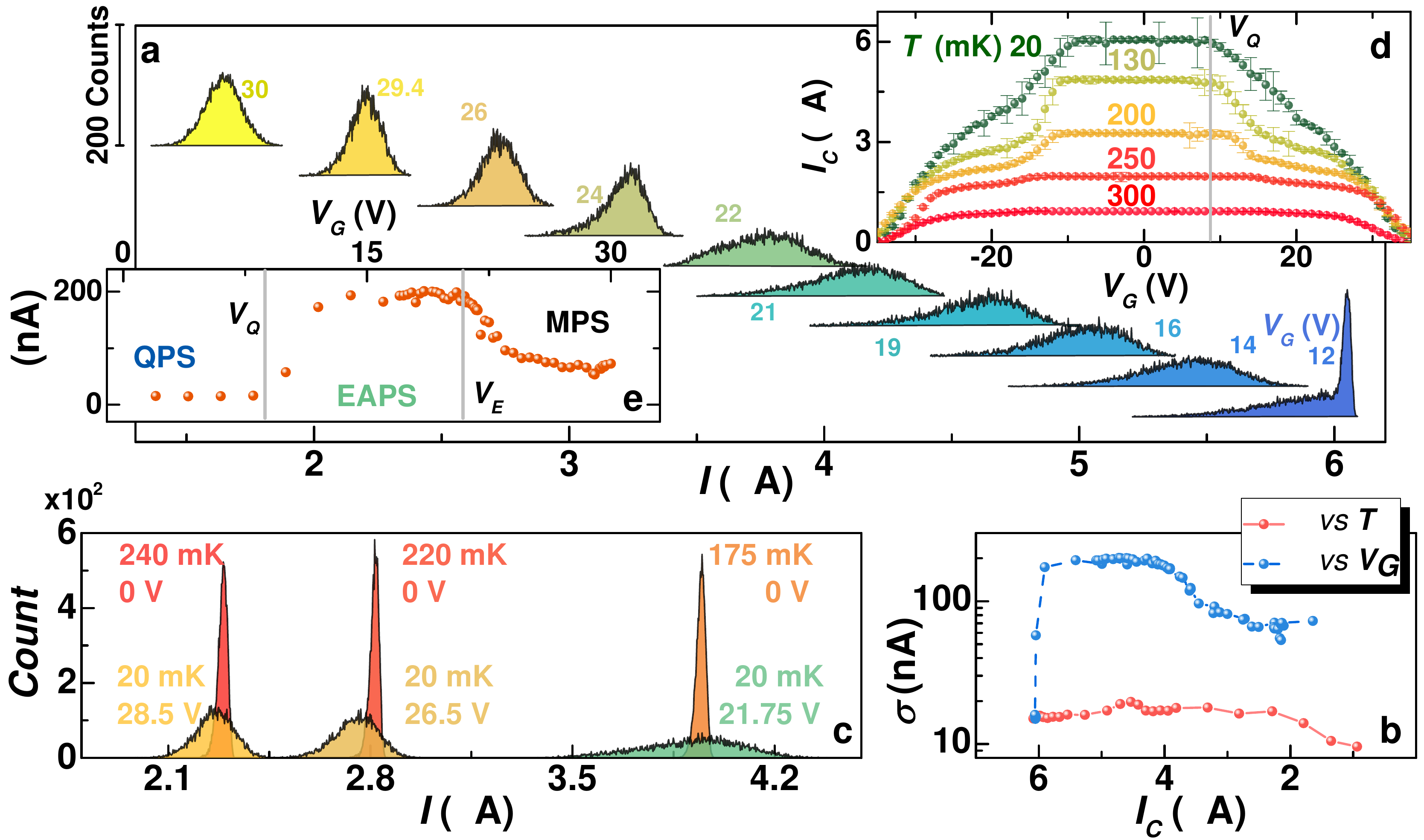}
\caption{(a) Switching current probability distributions vs current $I$ at different gate voltage values from 12V to 30V. 
The curves are vertically offset for clarity. (b) Comparison of $\sigma$ vs $I_C$  obtained for SCPDs as a function of  temperature and $V_G=0$ (light red), and of gate voltage at 20 mK (light blue). (c) Mode-matched SCPDs, red and orange distributions were obtained for $V_G=0$ at selected temperatures whereas yellow and green distributions were measured at $T=20$mK for different gate voltage values. (d) Dependence of the switching current $I_C$ on $V_G$ for different values of bath temperature from 20mK to 300mK. Data were obtained from the average computed over 25 acquisitions, and the error bars represent the standard deviation. The crossover voltage $V_Q\simeq8V$ separates the QPS from the EAPS regime. (e) Standard deviation $\sigma$ of the SCPDs vs  $V_G$. Crossover voltages $V_Q\simeq8V$ and $V_E\simeq21V$ separate QPS/EAPS and EAPS/MPS regimes, respectively.}
\label{fig:fig3}
\end{figure*}

\subsection{SCPDs measurements} 
The stochastic behavior of the switching current of a Dayem bridge can be modelled with the resistively and capacitively shunted junction (RCSJ) theory \cite{mccumber1968,stewart1968} which schematises a Josephson junction as the parallel of a resistor, a capacitor and a phase-dependent current generator $I(\phi)$. According to this model, we can interpret the transition to the normal state as a phase particle moving in a tilted washboard potential under the effect of a friction force (see Fig. \ref{fig:fig2}a) \cite{bezryadin2012}. In this framework, switching events are represented by the escape of the phase particle from a minimum of the potential, corresponding to a $2\pi$-rotation in $\phi$. 
The probability distribution $P(I,T)$ for this event to occur, as a function of the bias current $I$ and of the electronic temperature $T$, is given by the inverse Kurkijärvi–Fulton–Dunkleberger (KFD) transform \cite{fulton1971,Kurkijarvi1972}:
\begin{align*}
    P(I,T)=\frac{\Gamma(I,T)}{\nu_I}\exp{\left[-\frac{1}{\nu_I}\int_0^I\Gamma(I',T)dI'\right]},
    \label{eq:invkfd}
\end{align*}
where
\begin{align*}
\Gamma(I,T) = \frac{L}{2\pi \xi(T)\tau_{GL}(T)}\sqrt{\frac{\Delta U(I,T)}{k_B T}} \times
\\
\exp{\left[-\frac{\Delta U(I,T)}{k_B T}\right]}
\end{align*}
is the phase slip rate, $\nu_I=dI/dt$ is the ramp speed of the bias current, $\xi(T)$ is the Ginzburg-Landau (GL) coherence length, $\tau_{GL}(T)$ is the so-called GL relaxation time \cite{langer1967}, $\Delta U (I,T)=aE_J(T)\left(1-\frac{I}{I_C(T)}\right)^b$ is the height of the potential barrier, $E_J(T)=\hbar I_C(T)/2e$ is the Josephson energy, $e$ is the electron charge,  and ($a,b$) are parameters accounting for the typology of the Josephson weak-link \cite{bezryadin2012}.  
Conventionally, the critical current $I_C$ of the superconducting junction is assumed to be either the maximum current for which $P(I,T)\neq0$ or the mode of the SCPD; in the following we will adopt the latter definition. 

\subsubsection{Temperature dependence of SCPDs}
Figure \ref{fig:fig2}(b) shows the SCPDs built through $10^4$ acquisitions of the switching current measured at several temperatures ranging from 20 mK to 300 mK. 
To perform SCPD measurements we used a 750-KHz bandwidth input/output analog-to-digital/digital-to-analog converter (ADC/DAC) board for the acquisition of the voltage drop signal and the generation of the bias current, respectively.
The input signal consisted of an 8.7 Hz saw-tooth  current wave obtained by applying a voltage signal generated by the digital  board to an 1M$\,\Omega$ load resistor. The current wave was composed by a positive linear ramp with amplitude $10\,\mu$A, and slope $\nu_I=133\,\mu$A/s followed by a 100 ms zero-current plateau which turned out to be essential for the weak-link to cool down between two consecutive transitions to the normal state. 
In particular, the mode of the distributions decreases by raising the temperature, as a consequence of the reduction of $I_C$. The width and the shape of the distributions follow the conventional behaviour\cite{bezryadin2012,mccumber1970} as a function of the $T$, quantitatively described by the evolution of the  standard deviation $\sigma$. Firstly, the quantum phase slips (QPS) regime occurs when the transition from one minimum of the tilted washboard  potential to the next one is due to quantum tunneling. 
Since the tunneling process does not require an activation energy, the standard deviation of the SCPDs in this regime is expected to be temperature independent. 
The temperature range where tunneling is the main source of phase slips defines the so-called crossover temperature $T_Q$, above which $(T>T_Q)$ the thermal energy of the system allows the phase particle to hop over the potential barrier. 
In such thermally-activated phase slip (TAPS) regime, the thermal energy supplied to the system growths with the temperature, resulting into a widening of $\sigma$ as a function of $T$.
Finally, when thermal energy is large enough to allow more than one phase slip event to occur simultaneously ($T>T_{T}$), the system falls in the so-called thermally-activated multiple phase slips (MPS) regime, and the standard deviation is known to decrease as a function of $T$ \cite{bezryadin2012}. 

 The plot of $\sigma$ vs $T$ for the same Josephson nanotransistor, shown in Fig. \ref{fig:fig2}d, demonstrates that our Ti Dayem bridge follows the RCSJ model and the conventional phase slip theory \cite{giordano1988,giordano1989,kramers1940,bezryadin2012}. Indeed, from 20 mK up to 110 mK it shows an almost constant value of $\sigma$ of $\sim 15$ nA (QPS regime), from 110 mK to $\sim 150$ mK the standard deviation  is proportional to the temperature (TAPS regime), and for $T\gtrsim 150$ mK $\sigma$ decreases down to $\sim10$ nA at 300 mK (MPS regime). 
 We wish to stress that the independence of $\sigma$ in the QPS regime (i.e., $T\lesssim 110$ mK) cannot be ascribed to a saturation of the electronic temperature in the Josephson junction since the critical current turns out to increase in this range by decreasing $T$ (see Fig. \ref{fig:fig1}c).
 Both in the QPS and TAPS regimes it is possible to fit the SCPDs curves through the KFD transform\cite{Kurkijarvi1972,fulton1974,bezryadin2012} in order to extract the characteristic parameters of our system. 
 Dotted lines in Fig. \ref{fig:fig2}b represent the inverse KFD transform fits of our data which show good agreement with the theory. Fit parameters, and a more detailed description about the fitting procedure are provided in the Supplemental Material (SM).

Figure \ref{fig:fig2}c shows the JJ escape rate $\Gamma (I,T)$ computed through the direct KFD transform \cite{fulton1974,Kurkijarvi1972}
\begin{equation*}
    \Gamma (I_N,T)=\frac{P(I_N,T) \nu_I}{1-w\sum_{k=0}^N P(I_k,T)},
\end{equation*}
where $w$ is the bin size of the $P(I,T)$ histograms, and $P(I_k,T)$ is the switching probability in the current interval $[k w, (k+1)w]$ with $k\in \mathbb{N}$. 
$\Gamma(I_N,T)$ provides a measure of the phase lifetime of our Dayem bridges, which spans between 1 $\mu$s ($\Gamma\sim10^6$ Hz) and 10 ms ($\Gamma\sim 10^2$ Hz). 
The above escape rate range is in agreement with conventional switching current experiments performed so far \cite{bezryadin2012,bezryadin2000}.

\begin{figure*}[ht!]
\centering
\includegraphics[width=\linewidth]{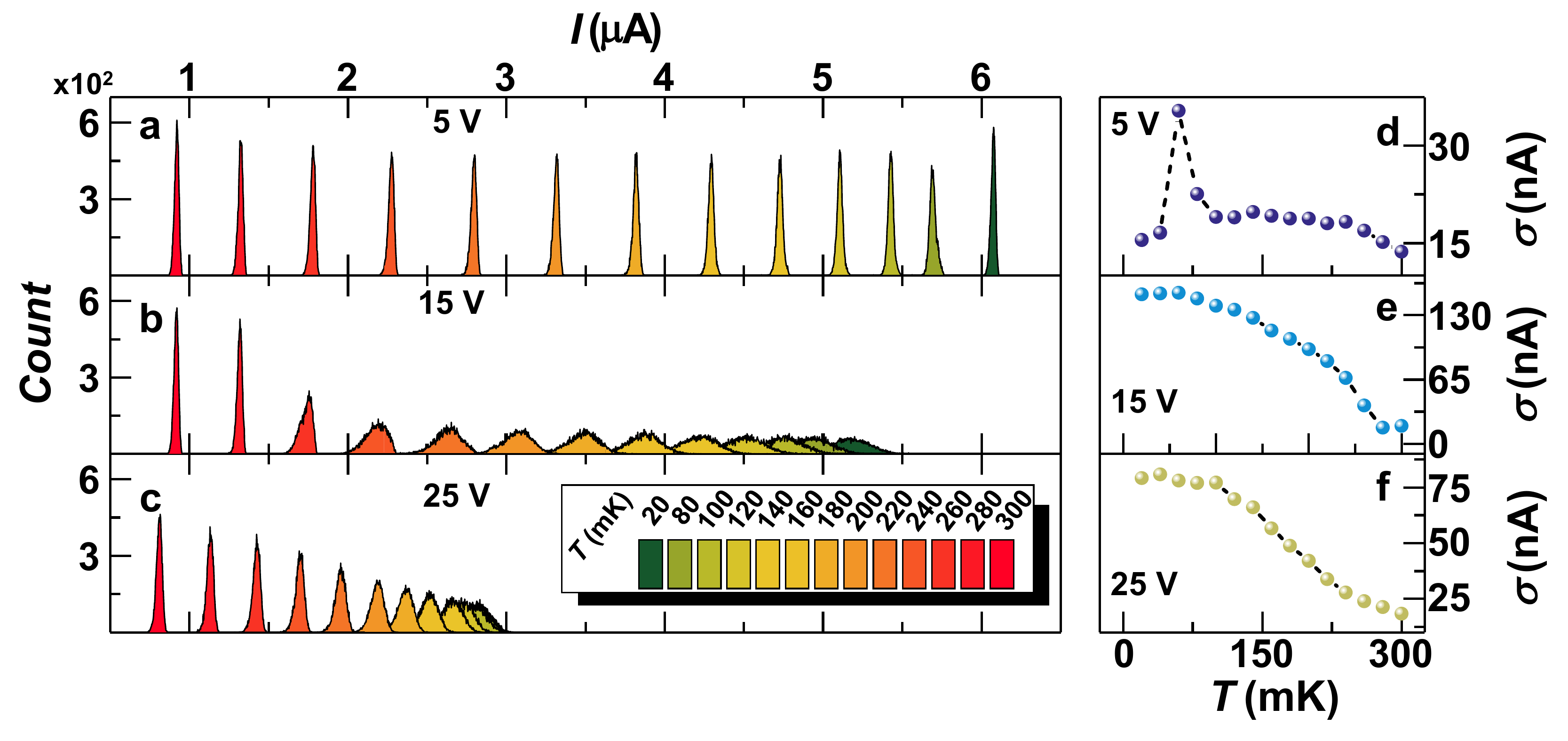}
\caption{(a,b,c) Evolution of the SCPDs from 20 mk to 300 mK for selected values of the gate voltage ($V_G=5,\ 15,\ 25$ V). Widening of the SCPDs caused by the electric field is clearly visible. 
(d,e,f) Dependence of standard deviation $\sigma$ on temperature $T$ for the SCPDs shown in panels (a,b,c). The Josephson weak-link turns out to  evolve towards the MPSs regime as the electric field increases.}
\label{fig:fig4}
\end{figure*}

\subsubsection{Impact of the electrostatic field on SCPDs}
Let us now focus on the characterization of the impact of the electrostatic field on the  phase slips  dynamics. 
To verify the customary\cite{desimoni2018,paolucci2018,paolucci2019,desimoni2019,Paolucci2019a} dependence of $I_C$ on the electric field, a detailed measurement of $I_C(V_G)$ as a function of bath temperature was preliminary performed on our Dayem bridge (see Fig. \ref{fig:fig3}d). The curves turns out to be symmetric for $V_G \rightarrow -V_G$, and show the expected monotonic suppression of the switching current, with full quenching at $|V_G^C|\simeq34$ V. Moreover, as the temperature grows, we observe the typical \cite{desimoni2018,paolucci2018,paolucci2019,desimoni2019,Paolucci2019a} plateau widening for low $V_G$ values. 
The current $(I_L)$ flowing between the constriction and the gate electrode was also measured, and it was found to be at most $I_L^{MAX}\simeq 15$ pA for $V_G=34$ V, corresponding to a gate-bridge resistance  $ R_L\simeq 2.3$ T$\Omega$. 
For a more comprehensive discussion on the effect of the gate current see the SM.

In order to asses the impact of the electric field on the SCPDs, we acquired the distributions at 20 mK for several different values of $V_G$. 
First of all, we emphasize that the application of an electric field to the weak-link dramatically deforms the shape of the SCPDs. 
In particular, as shown in Fig. \ref{fig:fig3}a, for $V_G<8$ V the SCPDs cannot be distinguished from the zero-gate one whereas
a tail at low current values appears for $8<V_G<14$ V. 
Moreover, the SCPDs strongly widens for 14 V $<V_G<$24 V. 
Finally, for $V_G>24$ V, the SCPDs turn out to narrow. The above behaviour is quantitatively described by the standard deviation of the distributions displayed in Fig.  \ref{fig:fig3}e.  Here, we note that in the $\sigma$ vs $V_G$ curve it can be still identified a region of constant standard deviation, thereby indicating a negligible contribution of the electric field to the phase slips for low $V_G$ values. 
This behavior turns out to be equivalent to the QPS regime. 
Notably, such a regime occurs in the voltage range (i.e., $|V_G|<V_Q$, see Figs. \ref{fig:fig3} d and e) where not even $I_C$ is affected by the electrostatic field. 
For $|V_G|>V_Q$, $I_C$ starts to monotonically decrease,
phase slips are activated by the application of the electrostatic field, and $\sigma$ grows with $V_G$ obtaining its maximum value of $\sim200$ nA. 
We define this region as the "electrically-activated" phase slip (EAPS) regime.
This evidence suggests that, whatever the microscopic origin of $I_C$ suppression, the latter is accompanied by a corresponding increase of phase slip events.

Finally, for higher values of the electric field (i.e., $|V_G|>V_E\sim20$ V), $\sigma$ decreases and saturates to $\sim 75$ nA, a value which is around 7.5 times larger than the corresponding one in the high temperature  case. 
Therefore, this behaviour, yet resembling the thermally-activated MPS regime, \emph{cannot} be ascribed to a conventional thermal trigger of  phase slips. To emphasize this point,  we compare the evolution of $\sigma(V_G,T=20$ mK) and $\sigma(V_G=0,T)$ by plotting them vs their corresponding critical current values (see Fig. \ref{fig:fig3}b).
We speculate that the electric field effect is responsible for a deep modification of the weak-link phase dynamics by enhancing the switching probability (i.e., fluctuations) in a wider current bias range. 
In addition, although both $\sigma$ vs $T$ and $\sigma$ vs $V_G$ curves present a similar behaviour, and three qualitatively-similar regimes are recognisable in either curves, the average value of $\sigma (V_G)$ is around one order of magnitude larger than that of $\sigma (T)$. 

To allow a further comparison between  thermal and electric field distributions, we  plot selected SCPDs corresponding to roughly the same $I_C$. Figure \ref{fig:fig3}c shows such mode-matched distributions for $I_C=2.2,\ 2.8,\ 4.0\ \mu$A.
The $I_C$-matched distributions show markedly different shapes and widths, a behavior which might stem from electrostatically-driven strong \emph{nonequilibrium}  induced in the superconducting  Dayem bridge. 
Yet, this provides an additional confirmation that electrostatic field effect cannot be explained by a trivial local  quasiparticle overheating of the superconductor. 
Indeed, assuming that the widening of the distributions were due to a thermal effect, the required effective quasiparticle temperature would be so high to be incompatible with the existence of superconductivity \cite{giazotto2006}. This observation reflects into a meaningless attempt to fit the electrically-activated SCPDs with a conventional KFD transform since the necessary parameters would be totally outside the range of validity.

It is finally noteworthy to examine the response of the Josephson bridge under the simultaneous action of an electric field and  thermal excitations. 
Figure \ref{fig:fig4}a,b,c show the SCPDs as a function of temperature in the range between 20 mK and 300 mK for $V_G=5,\ 15,\ 25\ V$, respectively. 
At high temperature, the distributions seem to recover the thermal behaviour for each value of the applied electric field. 
Such an effect demonstrates a weakening of the electric  field impact on phase slips at high temperature, which is consistent with what already observed in previous experiments \cite{paolucci2018,desimoni2018,paolucci2019,desimoni2019}, and in our preliminary electric field characterization of the critical current (see Fig. \ref{fig:fig3}d).  
Nonetheless, the evolution of $\sigma$ vs $T$  drastically changes when the electric field is applied. 
Figure \ref{fig:fig4}d,e,f show the evolution of $\sigma$ as a function of temperature for $V_G=5,\ 15,\ 25\ $ V. 
At low values of the gate voltage (i.e., $V_G=5$ V) we can identify  QPS, TAPS, and MPS regimes \cite{Li2011}. 
By contrast, for $V_G\geq15V$, the electric field seems to drive permanently the JJ into a regime that is qualitatively similar to the thermal MPS regime for every temperature value. 
In such a configuration, QPSs and TAPSs regimes cannot be observed anymore.

\section{Conclusions}
In conclusion, we have shown the occurrence of different phase slips regimes in a Ti Dayem bridge Josephson weak-link at several temperatures down to 20 mK. 
Firstly, the SCPDs of the system show the typical behaviour as a function of temperature. 
Secondly, the distribution shape is largely affected by an externally-applied  electrostatic field. In particular, the standard deviation of the SCPDs  far increases when the electric field is present. 
The drastic difference observed between the effect of the electric field and the temperature on SCPDs is a clear evidence of the \emph{non-thermal} origin of the field effect-driven critical current suppression, and could be ascribed to a strong nonequilibrium condition set in the weak-link. 
Finally, as far as the specific applicative interest of the switching current probability distribution measurements presented in our manuscript is concerned, we stress that, especially in view of the possible realization of superconducting field effect-based qubits (such as, \textit{e.g.} all-metallic gatemons) a good control and knowledge of phase noise and fluctuation sources in gated superconducting transistors is required. We conclude by highlighting that, in some sense, in our devices the gate voltage acts as an on-demand source of phase slippage in the Dayem bridge, which could be exploited to implement field effect-based platforms of phase-slip qubits (see Ref. \cite{mooij2005}).


\section*{Acknowledgement}
We acknowledge A. Braggio, R. Citro, V. Golovach, S. Kafanov, N. Ligato, Y. Pashkin, F. Paolucci, P. Solinas, and E. Strambini for fruitful discussions. The authors acknowledge the European Research Council under the European Union's Seventh Framework Programme (COMANCHE; European Research Council Grant No. 615187) and Horizon 2020 research and innovation program under Grant Agreement No.  800923 (SUPERTED) for partial financial support.

\section*{Appendix}

\subsection*{\NoCaseChange{Gate-Dayem bridge current}}
\begin{figure}[hb!]
	\centering
	\includegraphics[width=8cm]{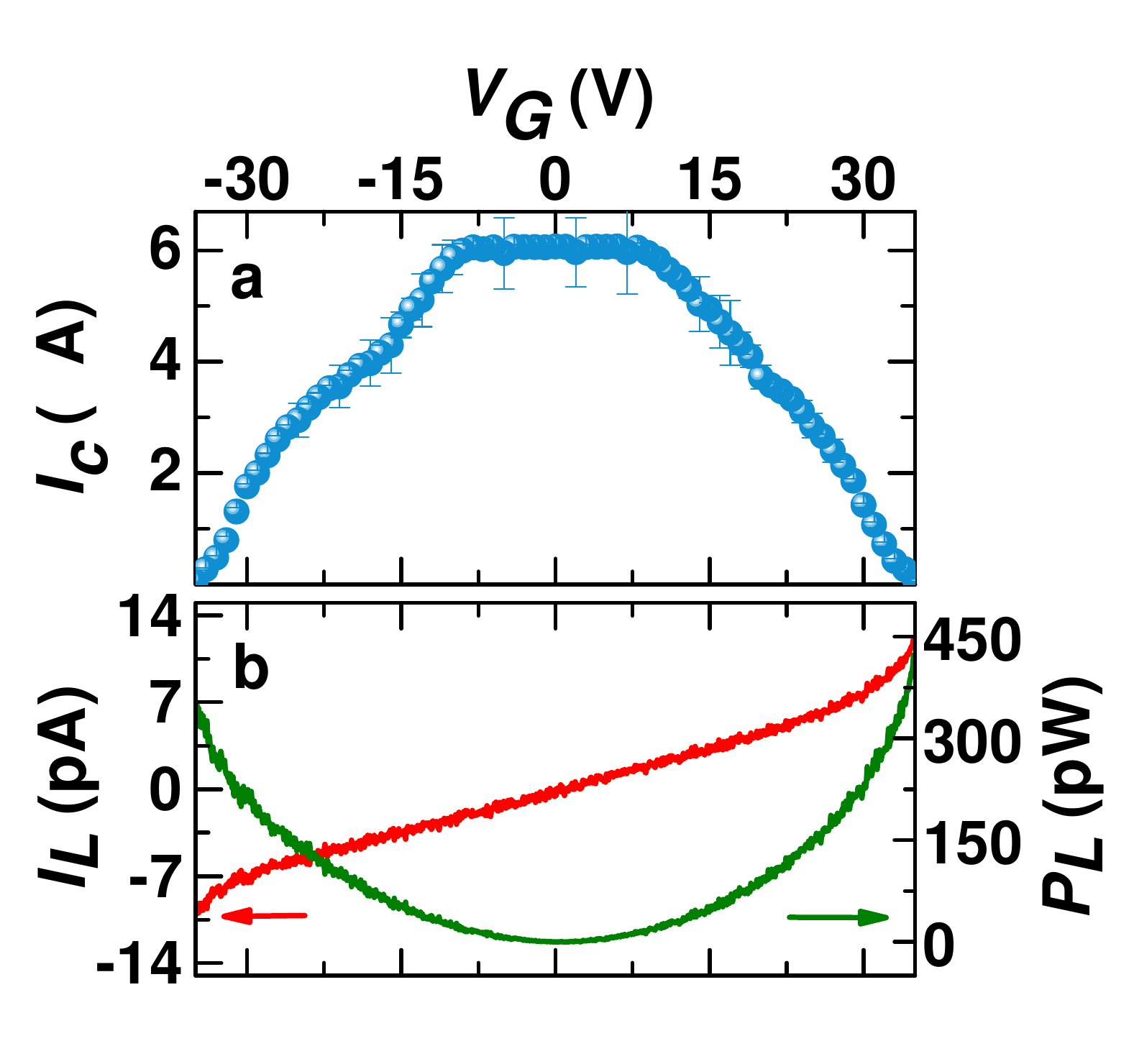}
	\caption{a) $I_C$ vs $V_G$ curve at a temperature of 20 mK for the same bridge device shown in the main text. Data are computed over 25 measure repetitions. b) Leakage current $I_L$ (orange line) and the corresponding Joule power $P_L$ (green line) as a function of the gate voltage $V_G$  measured at $T=20$ mK. The resulting gate-channel resistance is $R_G= 2.54$ T$\Omega$.}
	\label{fig:s1}
\end{figure}

The current between the Dayem bridge (DB) constriction and the gate was measured by a standard two-probes technique with a low noise voltage generator and a room-temperature current pre-amplifier. Figure \ref{fig:s1}a shows the $I_S$ vs $V_G$ characteristic for $T=20$ mK from -35 V to 35 V. The corresponding gate-Dayem bridge current $I_L$ (See Fig. \ref{fig:s1}b) as a function of $V_G$ displays, in agreement with the conventional theory for electron tunnel-injection at low biases \cite{simmons1963}, a linear behaviour for almost the entire explored range,  with a maximum value of $I_L^{MAX}=12.5$ pA at $V_G=35$ V which is on par with previous similar experiments \cite{desimoni2018,paolucci2018,paolucci2019,desimoni2019,Paolucci2019a}.  
This measurement provides only an upper boundary for the current injected into or extracted from the DB: indeed, even neglecting leakages in the electrical setup lines, a fraction of $I_L$ is expected to directly reach the leads. The green line in Fig. \ref{fig:s1}b represents the total power injected into the system. Its maximum, $P_L=400$ pW, occurs for $|V_G|=35$ V. We emphasize that such a power is unlikely to be directly dissipated into the weak link: the current between the gate and the DB can be described by the parallel of a possible diffusive current through the substrate and of a ballistic transport in the vacuum due to field emission at high electric fields, if present.
\newpage
To correctly assess the impact of the diffusive current, it is necessary to take into account for the ratio $N$ between the length of the shortest diffusive path connecting the gate and the DB (d$\sim$ 50 nm) and the average scattering length in sapphire substrates ($\lambda_s\lesssim$ 0.1 nm, see \cite{Hughes1979}), providing the average number of collision of carriers with phonons and defects. We stress that, in such semiclassical approach, after each scattering event, the carrier motion is randomized. After that, it accelerates in the electric field, until it scatters again. Provided that in our devices $N\sim500$, electrons and holes  reach the Dayem bridge after having relaxed their kinetic energy several times. This suggests that such carriers are unlikely to have a role in raising the electronic temperature of the DB. Therefore, we believe that the diffusive current in the substrate cannot be taken as the origin the observed phenomenology.
	
We now focus on the process for which an electron is field-emitted from the gate and absorbed by the Dayem bridge. An electron with an energy between 1 and 30 eV, and ballistically reaching the junction through the vacuum, releases its energy causing an abrupt increase of the electronic temperature. \cite{giazotto2006,Timofeev2009}. The electronic contribution to the heat capacitance $C_e$ in a weak link in the normal state is:
	$$C_e=\Omega\cdot \gamma\cdot T_e,$$
	where $\Omega$ is the volume of the junction, $\gamma=\pi^2 k_B^2 \nu_F/3$ is the Sommerfeld constant, $\nu_F=1.35\times10^{47}$ m$^{-3}$J$^{-1}$ is the density of states at the Fermi level for Ti, and  $T_e$ is the electronic temperature of the system. The released energy $E(V)$ is proportional to the acceleration voltage ($V$) between the gate electrode and the Dayem bridge: 
	$$E(V)=q\cdot V,$$             $$ P(t)=E \delta(t),$$
	where $q$ is the electron charge, $\delta$ is Dirac delta and $P(t)$ is the impulse power as a function of the time $t$. According to heat transport theory, the evolution of the electronic temperature in the junction is described by the following differential equation (see see Ref. \cite{giazotto2006}), where $T_B$ is the lattice temperature:
	$$C_e  \frac{\partial T_e}{\partial t}=P(t),$$  
	$$T_e=\sqrt {\frac{2 E}{\Omega \gamma}+T_B^2}\sim10 K.$$
	We wish to stress that $T_e$ is an underestimate of the final electronic temperature of the weak link because we assumed $C_e$ to be that of the normal state, which is typically exponentially larger than in the superconducting state owing to the presence of the energy gap in the density of states. 
	The above calculation shows that a single electron with an energy about $E=30$ eV injected into the Dayem bridge at $T_B=10$ mK would raise its electronic temperature up to a value which is more than 20 times larger than its critical temperature ($T_C\simeq300$ mK).  This result allows us to make the statement that the heat originating from field-emitted electron absorption in the bridge cannot result into an equilibrium condition with a definite gate-controllable electronic temperature and critical current. Rather, we are forced to make the hypothesis that, due to continuous absorption of highly-energetic electrons, the Dayem bridge bounces continuously between its normal and superconducting states. If this were the case, every time an electron is absorbed, it suddenly makes the bridge resistive, which then relaxes back to the superconducting state. The periodicity of such events is given by the electron emission rate, while the relaxation time is essentially given by the electron-phonon relaxation time, which is expected to be of the order of 1 ns  (see Ref. \cite{giazotto2006})), i.e., much lower than the typical integration time of our  measurement setup ($\simeq20$ ms). In this scenario, during an $I-V$ measurement, every time an electron is absorbed by the Dayem bridge when $I$ is below the retrapping current ($I_R$) the variation of its resistance is expected to be so fast to be undetectable with our setup. By contrast, when $I>I_R$, each time an electron is absorbed by the Dayem bridge, the latter should immediately switch to the normal state, and should persist in such condition until the bias current is set back to 0. This implies that when field emission occurs the retrapping and the switching current should always coincide. Since this is not the case, we believe that we should exclude any hot electron injection mechanism related to field emission as the predominant origin of our observations.
\begin{figure*}[b!]
\label{s2}
\centering
\includegraphics[width=\textwidth]{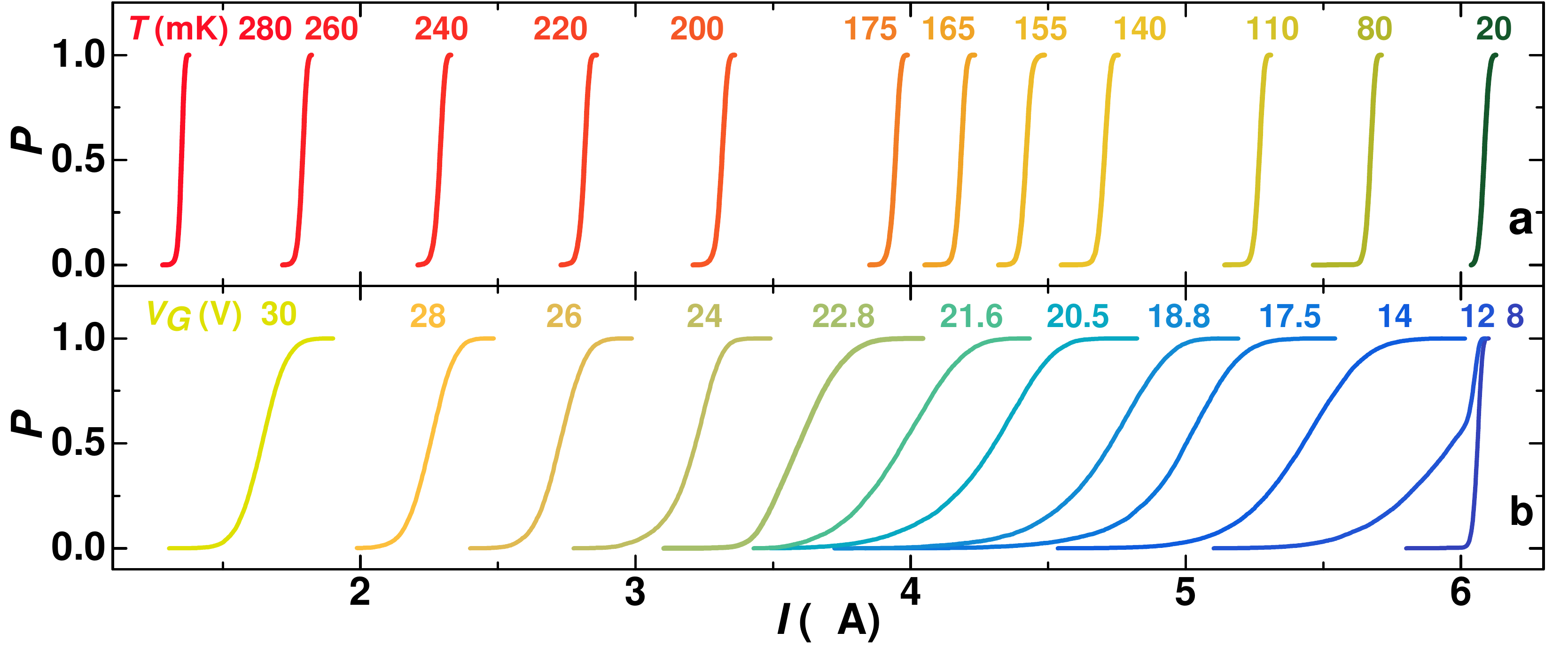}
\caption{a) SCCPDs (S-curves) as a function of the temperature from 20 mK to 280 mK. b) S-curves for different values of the gate voltage $V_G$ from 8 V to 30 V.}
\end{figure*}
\subsection*{\NoCaseChange{Inverse KFD transform fit}}
\begin{figure*}[ht!]
\centering
\includegraphics[width=\textwidth]{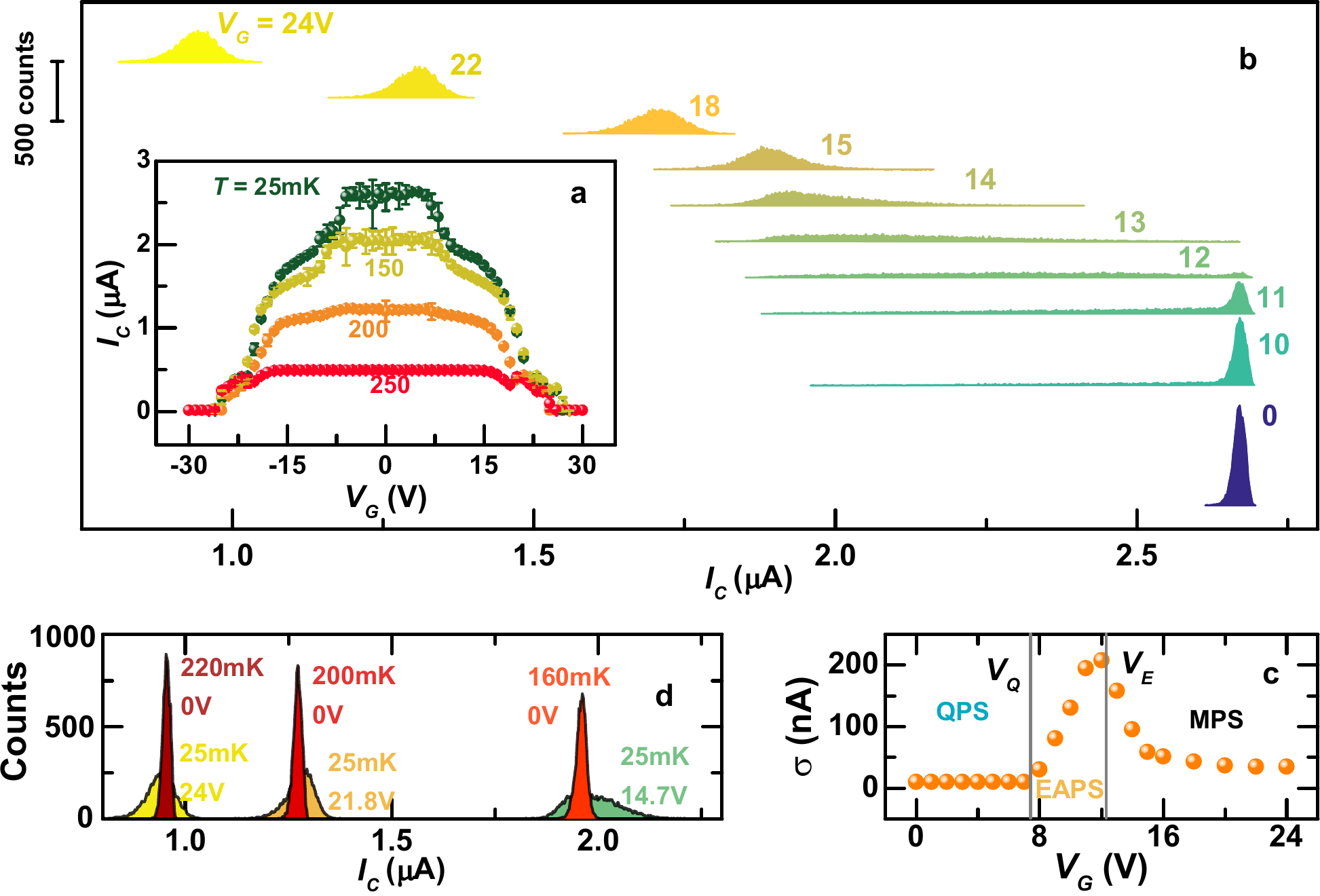}
\caption{a) Dependence of the switching current $I_C$ on $V_G$ for different values of bath temperature from 25 mK to 250 mK. Data were obtained from the average computed over 25 acquisitions, and the error bars represent the standard deviation.
 b) Switching current probability distributions vs current $I$ at different gate voltage values from 0 V to 24 V. 
The curves are vertically offset for clarity.
c) Standard deviation $\sigma$ of the SCPDs vs  $V_G$. Crossover voltages $V_Q\simeq8V$ and $V_E\simeq12V$ separate QPS/EAPS and EAPS/MPS regimes, respectively. d) Mode-matched  SCPDs,  red  and  orange  distributions  were  obtained  for $V_G=  0V$ at selected temperatures whereas yellow and green distributions were measured at $T= 20$ mK for different gate voltage values.}
\label{s3}
\end{figure*}
The fit was performed with the inverse KFD transform \cite{fulton1974,Kurkijarvi1972,bezryadin2012}:
\begin{align*}
    P(I,T)=\frac{\Gamma(I,T)}{\nu_I}\exp{\left[-\frac{1}{\nu_I}\int_0^I\Gamma(I',T)dI'\right]}
\end{align*}
where $\nu_I=dI/dt$ is the ramp speed of the bias current, and $\Gamma(I,T)$ is the phase slip rate which assumes the following expression for the TAPS and the QPS regime, respectively:
\begin{align*}
    \Gamma_{TAPS}(I,T)=\frac{L}{2\pi \xi(T)\tau_{GL}(T)}\times
    \\
    \sqrt{-\frac{a E_J(T)}{k_B T}\left(1-\frac{I}{I_C(T)}\right)^b}\exp{\left[-\frac{a E_J(T)}{k_B T}\left(1-\frac{I}{I_C(T)}\right)^b\right]},
\end{align*}
\begin{align*}
    \Gamma_{QPS}(I,T,T_{QPS})=\frac{L}{2\pi \xi(T)\tau_{GL}(T)}\times\\
    \sqrt{-\frac{a E_J(T)}{k_B T_{QPS}}\left(1-\frac{I}{I_C(T)}\right)^b}
    \exp{\left[-\frac{a E_J(T)}{k_B T_{QPS}}\left(1-\frac{I}{I_C(T)}\right)^b\right]},
\end{align*}
where $L$ is the geometric length of the weak link, $\xi(T)$ is the Ginzburg-Landau (GL) coherence length, $\tau_{GL}(T)$ is the GL time constant, $E_J$ is the Josephson energy, and $a$ and $b$ are parameters accounting for the typology of the Josephson weak-link. Their values can be analytically derived for tunnel Josephson junctions $(a_{tunnel}=4\sqrt{2}/3\ ,\ b_{tunnel}=3/2)$ \cite{bezryadin2012} and for long \emph{metallic} superconducting wires $(a_{LW}=\sqrt{6}\ ,\ b_{LW}=5/4)$ \cite{sahu2009}, but are not known for Dayem bridges, therefore they are left as free parameters of the fit. For all the fit of SCPD curves, $a_{DB}$ and $b_{DB}$ converged to the same values $a_{DB}=1.0\pm0.1$ and $b_{DB}=1.40\pm0.01$ Also, in the QPS regime (20 mK $\leq$ T  $\leq$ 90 mK) we introduced the effective temperature $T_{QPS}$ as a fitting parameter.
\begin{table}[h!]
\label{table1}
\centering
\begin{tabular}{|l|l|l|l|}
\hline
\hline
$T$ (K)  & $T_{QPS}$ (mK) \\
\hline
\hline
0.02 & $99\pm4$  \\
\hline
0.03 & 110 $\pm5$    \\
\hline
0.04  & $114\pm7$ \\
\hline
0.05  & $115\pm 6$   \\
\hline
0.06  & $119\pm7 $   \\
\hline
0.07  & $125\pm9 $   \\
\hline
0.08  & $135\pm7$    \\
\hline
0.09  & $153\pm8 $  \\
\hline
\hline
\end{tabular}
\caption{$T_{QPS}$ values yielded by the fitting procedures of the SCPDs with the inverse KFD transform.}
\end{table}\\
The value for $T_{QPS}$ yielded by the fitting procedure are shown in the Table 1.
We note from the point of view of the Josephson coupling, our weak links are in-between a tunnel junction and a long metallic superconducting wire just like the value found for parameter $b_{DB}$ is in-between the other types of junctions $(b_{LW}<b_{DB}<b_{SIS})$. 

\subsection*{\NoCaseChange{Switching current cumulative probability distributions}}
The switching current cumulative probability distribution (SCCPD), or S-curve, is defined as the integral of the switching current probability distribution (SCPD) \cite{Zgirski2018}. In this work, the S-curves are obtained upon summation of the frequency counts of the SCPD histograms shown in the main text. The SCCPDs describe the probability to find the system in the normal state for a given value of the current bias.

Figure \ref{s2} shows the evolution of the SCCPD as a function of the temperature $T$ ($V_G=0$ V) and the gate voltage $V_G$ ($T=20$ mK). In particular, the application of the electric field results in a sizable widening of the S-curves. A comparison between S-curves with similar $I_C$ values obtained in the thermal excitation case and in the electrostatic case allows to appreciate how broader are the SCCPDs in the latter case. This fact is a further evidence that field effect cannot be ascribed to a conventional "thermal-like" quasiparticle overheating in the weak-link region.

\subsection*{\NoCaseChange{Characterization of a second Josephson Dayem bridge}}
In this section, we show the characterization of a Ti Dayem bridge weak-link  similar to the one described in the main text. Qualitatively, the results obtained on this device resembles those presented in the body of the  manuscript. 
The thermal investigation of the system confirmed the typical behaviour of this kind of weak links \cite{bezryadin2012,paolucci2018,mccumber1968,stewart1968}. The preliminary characterization of $I_S$ response to the electric field (see Fig. \ref{s3}a) shows the nearly-symmetric  suppression \cite{desimoni2018,desimoni2019,paolucci2019,Paolucci2019b,Paolucci2019a} for both positive and negative gate voltage values, with full quenching at $|V_G|\simeq24$ V.
More interesting is the evolution of the SCPDs as a function of the gate voltage $V_G$ displayed in Fig. \ref{s3}b. As already similarly shown in the main text, the electric field modifies dramatically the shape of the SCPDs following the behaviour described in the main body of the paper. Here, we note that the evolution of $\sigma$ as a function of $V_G$ (see Fig. \ref{s3}c) clearly displays as well the three regimes of quantum phase slips, electrical activated phase slips, and multiple phase slips. To compare the thermal and electric field distributions, we plot selected SCPDs with almost mode-matched $I_C$s. Figure \ref{s3}d shows such distributions for $I_C= 0.9,\ 1.2,\ 1.9\, \mu$A. The $I_C$-matched SCPDs show drastically different shapes and widths.


%

\end{document}